# $(Pb_{1-x}Bi_x)(Ti_{1-x}Mn_x)O_3$: Competing mechanism of Tetragonal-Cubic phase on A/B site modifications

Arun Kumar Yadav[1], Anita Verma[1], Birender Singh[4], Deepu Kumar[4], Sunil Kumar[1], Velaga Srihari[5], Himanshu K. Poshwal[5], Pradeep Kumar[4], Shun-Wei Liu[3], Sajal Biring[3*], Somaditya Sen[1,2,3*]

[1]Discipline of Metallurgy Engineering and Materials Science, Indian Institute of Technology Indore, Indore 453552, India

[2]Department of Physics, Indian Institute of Technology Indore, Indore 453552, India

[3]Electronic Engg., Ming Chi University of Technology, New Taipei City, Taiwan

[4]School of Basic Sciences, Indian Institute of Technology Mandi, Mandi-175005, India

[5]High Pressure & Synchrotron Radiation Physics Division, Bhabha Atomic Research Centre, 400085, Mumbai, India

**Abstract**

Structural, vibrational and dielectric properties of $(Pb_{1-x}Bi_x)(Ti_{1-x}Mn_x)O_3$ (PBTM) ($0 \leq x \leq 0.50$) polycrystalline ceramics have been examined as a function of temperature. Synchrotron-based powder x-ray diffraction was employed to confirm phase purity and crystal structure of samples. Tetragonality ($c/a$ ratio) of the PBTM system exhibit an increase 1.065, 1.066 for $x = 0.06$ and 0.09 compositions respectively, compare to 1.064 for $x = 0$ sample. Curie point was found ~763 K and 773 K for $x = 0.06$ and 0.09 samples respectively. Though, tetragonality start to decreasing from samples with $x \geq 0.18$, Curie point of the samples started decreasing for $x \geq 0.12$. Temperature-dependent x-ray diffraction reveals the structural change from tetragonal to cubic structure. Unit cell volume was found to decrease with increasing temperature; indicate these materials are negative thermal expansion type until phase transition temperature, with negative thermal coefficients are $-1.571*10^{-5}$ /K, $-2.44*10^{-5}$ /K, and $-5.025*10^{-6}$ /K for $x = 0$, 0.06, and 0.09 samples respectively. Raman spectra showed softening of the transverse optical phonon modes and increase in the full width at half maxima with the increase in composition. Field-emission scanning electron microscope equipped with energy dispersive x-ray spectrometer (EDS) confirmed compositional homogeneity and dense type microstructure.

## Introduction

Structure-property relationships are significantly important in material physics. A good understanding of such relationships can be utilized to develop novel materials with enhanced qualities. $ABO_3$ type perovskite-based ferroelectric materials have been at the centre of the scientific and technological community because of their interesting structure. These materials have attracted much attention due to their enormous applications[1-3]. $BaTiO_3$, $PbTiO_3$, and $KNbO_3$ based materials are known to be excellent ferroelectric materials. PZT ($PbTi_{(1-x)}Zr_xO_3$) is one of the best industrialized ferroelectric/piezoelectric materials[4, 5]. Stability of ferroelectric properties in $BaTiO_3$ and $PbTiO_3$ systems are recognized due to condensation of a transverse optic zone-centre soft mode phonon. As a consequence of this, B-centre cations of oxygen octahedra take up off-centered positions. These systems are B-site governed ferroelectrics. Off-centering of B-sites increases bonding between their 'd' orbitals and the surrounding '2p' orbitals [2, 6]. At room temperature $PbTiO_3$ crystalizes in tetragonal structure ($C_{4v}^1$, *P4mm*, space group) and undergo structural transition to cubic perovskite ($O_h^1$, *Pm3m*, space group) at transition temperature $T_C \sim 490$ °C [7]. Piezoelectric materials have been widely used in many fields, due to its outstanding applications as transducers, sensors, actuators, etc. $PbTiO_3$-based ferroelectric ceramics have been extensively explored for many years because of their excellent ferroelectric/piezoelectric properties. However, restriction of environmental concerns due to toxic nature of Pb, and de-poling of PZT ceramics above 200 °C have forced to investigate for ferroelectric materials with reduced Pb content but with high Curie temperature[8, 9].

In a simple diatomic molecule, with the increase in thermal energy, bond length increases, thereby vibrational energy increases as well. In a typical structure, entire phonon density of states varies with increase in temperature. Enhancement in energies of vibrational modes with increase in temperature, give rise to positive thermal expansion. However, few categories of materials are there for which thermal expansion of chemical bonds may be affected by other factors which lead to shrinkage in volume. This type of materials in a certain temperature range show unusual property of negative thermal expansion behavior. There are many potential uses for materials with negative thermal expansion. They can be utilized with high precision technology such as optical mirrors, packaging materials for refractive index gratings, etc[10, 11]. Bi-based perovskites are some of the more prominent negative thermal expansion system. Off-centering in these materials occur due to chemical bonding is driven force of '6s$^2$' lone pair of Bi(6s$^2$) ions. Bi(6s$^2$) has more prominent off-centering capability than Pb(6s$^2$) based ions[12]. Some Bi-based ferroelectric materials are BiAlO$_3$, BiScO$_3$, Bi(Mg$_{0.5}$Ti$_{0.5}$)O$_3$, Bi(Mg$_{3/4}$W$_{1/4}$)O$_3$, Bi$_{0.5}$Na$_{0.5}$TiO$_3$, and Bi(Zn$_{1/2}$Ti$_{1/2}$)O$_3$, etc.[1, 10, 13-16]. Motivation of the present work is to elaborate the structural and electrical properties of (Pb$_{1-x}$Bi$_x$)(Ti$_{1-x}$Mn$_x$)O$_3$ ceramics. Sol-gel combustion method is used to synthesize these materials for $0 \leq x \leq 0.5$ range compositions. In this work, composition dependent structural and dielectric properties and interesting negative thermal expansion coefficients of some of these materials are studied.

**Synthesis and characterizations**

Polycrystalline (Pb$_{1-x}$Bi$_x$)(Ti$_{1-x}$Mn$_x$)O$_3$ ($0 \leq x \leq 0.50$) (named as PBTM) powders were synthesized using sol-gel combustion process. Precursors selected to synthesize these materials were Puratronic grade lead (II) nitrate, manganese nitrate (50 % w/w aqua solution), bismuth nitrate, and dihydroxy bis (ammonium lactate) titanium (IV) (50 % w/w aqua solution). Precursors were selected on the basis of solubility in doubly de-ionized (DI) water. Stoichiometric solutions of each precursor were prepared with DI water in separate glass beakers. Manganese solution was added to titanium solution. Bismuth and lead solutions were added to mixture subsequently. 5 % excess lead nitrate was used accounting for the volatile nature of *Pb* during sintering at high temperatures. The mixed solution was stirred for one hour to ensure proper homogeneity in the solution at normal temperature. A gel former was prepared in a separate beaker by mixing citric acid and ethylene glycol in 1:1 molar ratio in solution. This gel former solution was added in to homogeneous precursor solution. This solution containing constituent ions was vigorously stirred and heated (~80 °C) on a magnetic stirrer hotplate until a gel was formed. Gels were burnt inside a fume hood. Burnt powders were ground carefully in a mortar and pestle. Thereafter powders were heated at 500 °C for 12 h to get rid of trapped nitrates and polymers. These resultant powders were grounded again in mortar/pestle and re-heated at 700 °C for 12 h. After that, powders were mixed with 5 % PVA solution and pressed into pellets of 13 mm diameter and 1.5 mm thickness using a uniaxial hydraulic press. These pellets were heated first at 600 °C for 6 h to burn out binders and thereafter sintered continuously at 1000 °C for 6 h to form the dense pellets. Pellets were covered with powders of the same composition during heating to reduce Pb loss.

Synchrotron-based powder angle dispersive x-ray diffraction measurements (ADXRD) was carried out at Extreme Conditions Angle Dispersive/Energy dispersive x-ray diffraction (EC-AD/ED-XRD) beamline (BL-11) at Indus-2 synchrotron source, RRCAT, India, to confirm phase purity and structure of samples. From the white light of bending magnet, a wavelength of λ = 0.4457 Å was selected using a Si(111) channel-cut monochromator and then focused with a Kirkpatrick-Baez (K-B) mirror for ADXRD measurements. A MAR345 image plate detector (area detector) was used to collect two-dimensional diffraction data. Wavelength of the beam and sample to detector distance were calibrated using NIST standards $LaB_6$ sample. Calibration and conversion/integration of *2D* diffraction data to *1D*, intensity versus 2θ, was carried out using FIT2D software[17]. In order to reduce the preferred orientation effects, measurements were carried on crushed powders filled in capillary, which was rotated at ~150 rpm. The linear absorption coefficient of sample and 25% packing density of material in capillaries were kept in mind in selecting the diameter of the capillary tube. Temperature-dependent SRPXRD studies were performed for some selected samples using STOE high temperature attachment.

Room temperature micro-Raman measurements were carried out in backscattering geometry with a Labram HR-Evolution Raman spectrometer. A green 532 nm laser was used as excitation and focused on to the samples through a 50xLWD microscope objective. Scattered light dispersed through 1800 lines/mm grating was detected with Peltier cooled charged coupled device (CCD) detector. The laser power was kept low (~1 mW ) on the sample to avoid heating effects. Microstructure and grain size of the sintered pellets were investigated by Supra 55 Carl Zeiss field emission scanning electron microscope. For electrical measurement, electrodes were prepared using high-temperature silver paste on both sides of sintered pellets. The silver pasted pellets on either side were cured at 550 °C for 15 minutes to establish good pellet-electrode contact. Before measurements, samples were heated at 200 °C for 15 minutes to get rid of surface moisture. The dielectric response was measured using a Newton's 4[th] Ltd. PSM 1735 phase sensitive LCR meter with the signal strength of ~1V.

[4]

**Results and discussion**

    Ambient synchrotron-based powder x-ray diffraction (named as SRPXRD) patterns for $0 \leq x \leq 0.50$ are shown in Fig. 1a. SRPXRD patterns, of samples with $x \leq 0.37$, belong to a pure tetragonal phase with *P4mm* space group, whereas for $x = 0.50$ composition diffraction peaks corresponding to cubic *Pm3m* space group. Variations of peak intensity with the composition ($x$) in $0 \leq x \leq 0.50$ range are shown in the Fig. 1b, highlighted part of Fig. 1a. To confirm the actual changes in structure as well as the phase purity of the sample, we have performed the Rietveld refinement on the SRPXRD data.

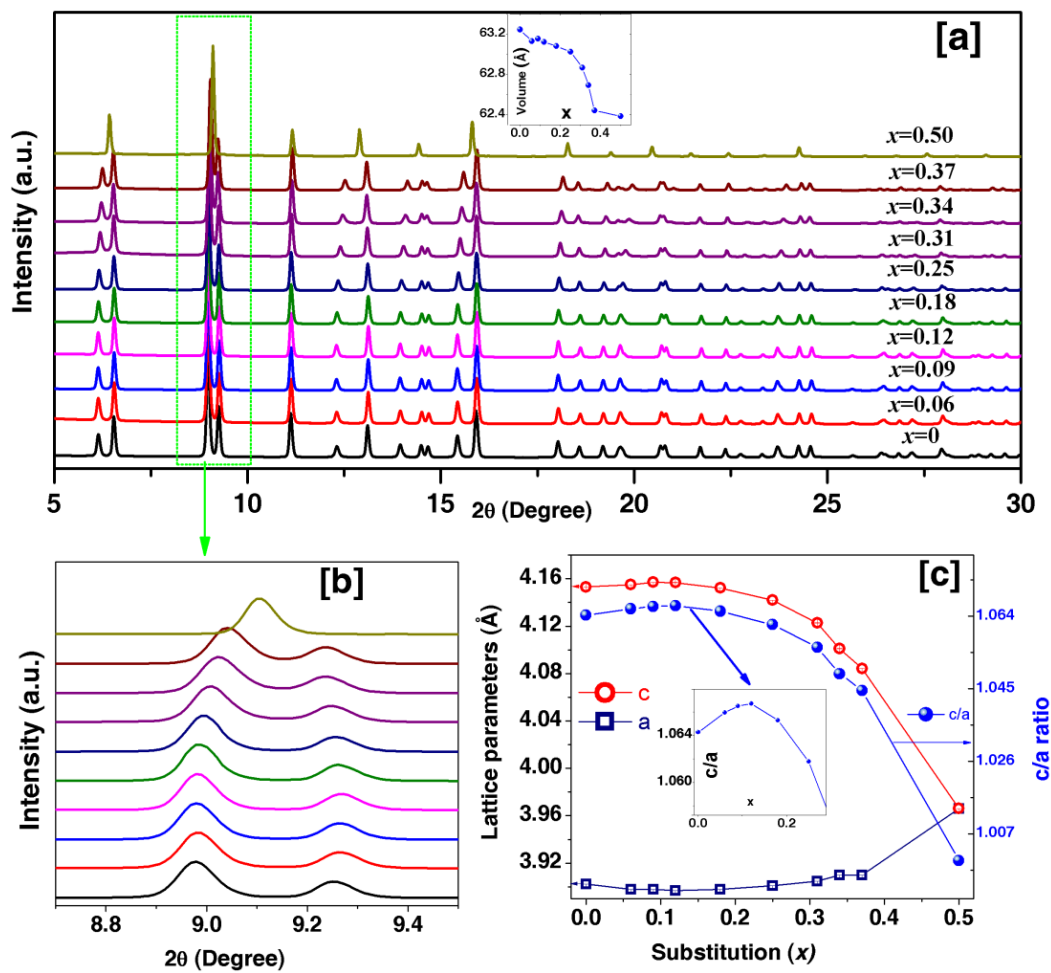

***Figure 1.*** (a) *SRPXRD of $(Pb_{1-x}Bi_x)(Ti_{1-x}Mn_x)O_3$ ($0 \leq x \leq 0.50$) range, calculated unit cell volume as a function of compositions are shown in inset fig. 1a., (b) Highlighted part x-ray spectra (c) Calculated lattice parameters and c/a ratio as a function of compositions, inset figure is showing the zoom part of the c/a ratio mark with arrow.*

The initial structural model for Rietveld refinement was taken as pure PbTiO$_3$ with the atomic positions being given in the non-centrosymmetric space group *P4mm* (No. 99) with Pb/Bi at the origin (0, 0, 0) [11, 18]. Background estimation was performed using linear interpolation between a set of background points and profile function was chosen as axial divergence asymmetric Pseudo-Voigt function. Parameters such as scale factor, lattice parameters, profile parameters (U, V, W) and position parameters (X, Y, Z) were refined one after the other. After successive refinement of these parameters, calculated and observed patterns show excellent match with the reasonable goodness of fit parameters ($\chi^2$, R$_p$, R$_{wp}$, R$_{exp}$). Representative plot with the goodness of fit parameter for all samples are shown in the supplementary information file (Fig. S1 and S2). The fitting R-factors are in acceptable range for all samples. A schematic representation of unit cell is shown in Fig. S3 (supplementary information file) for *x* = 0 and *x* = 0.50 compositions with the help of VESTA software [19]. Position parameters for all the samples obtained from Rietveld refinements are provided in the Supplementary information Table S1.

The calculated lattice parameters (*a*, *c*) with *c/a* ratio as a function of composition are shown in Fig.1c. Lattice parameters '*a*' slightly decreases while '*c*' increases upto *x* = 0.12 composition. Hence, tetragonality (*c/a* ratio) nominally increased initially in the range $0 \leq x \leq$ 0.12 from 1.064 to ~1.066, thereafter decreasing continuously in the range $0.12 \leq x \leq$ 0.50 to 1, for *x* = 0.50 (*c/a* = 1). At A-site, crystal radii of Bi$^{3+}$(XII) (~1.58 Å) is lesser than Pb$^{2+}$(XII) (1.63 Å) while for B-site Mn$^{3+}$(VI) (0.72 Å for low spin) and Mn$^{4+}$(VI) (0.67 Å) are less than Ti$^{4+}$(VI) (0.745 Å). Only Mn$^{3+}$(VI) (0.785 Å for high spin) is larger than the Ti$^{4+}$ ion [20]. Though, the crystal radii of substituent ions were smaller than the host ions, the observed unusual decrease in *a* and increase in *c* parameter at low concentration of *x* can only be explained with the chemistry associated with Bi-ion lone pair. Bi(6s$^2$) lone pair is more prominent than the Pb(6s$^2$), hence hybridization between Bi(6s$^2$) and O(2p) is more strong than Pb(6s$^2$) and O(2p). Upto a critical concentration of the Bi-ion in the system such as in this case $x \leq 0.12$, the ionic size effect is not only the controlling factor in the structural distortion. After that decrease of the lattice parameters are controlled by the ionic size effect at A and B-site [1]. However, the overall unit cell volume decreased with increase in substitution as shown in inset of Fig. 1a.

Temperature-dependent SRPXRD (HT-SRPXRD) ($\lambda$ = 0.5036 Å) was performed, in same geometry, to study phase transition temperature and lattice dynamics on selected samples (*x* = 0, 0.06, 0.09 compositions). Thermal equilibrium of samples was obtained by keeping the temperature constant for 10 min at the desired temperature, before start of experiment at each temperature. Non-centrosymmetric tetragonal to centrosymmetric cubic phase transformation is observed [Fig. S4 (a-c) and Fig. 2a] from merging of tetragonal peaks [Fig. S4 (b-d) and Fig. 2b] in the SRPXRD plots with the increase in temperature. Temperature-dependence of lattice parameters was estimated by refinement of HT-SRPXRD data. By using Le Bail profile fitting technique as implemented in Fullprof software to estimate lattice parameters of HT-SRPXRD data. Calculated lattice parameters for $x = 0$, 0.06, and 0.09 samples are shown in Fig. 2c as a function of temperature. Lattice parameters '$a$' is increasing while '$c$' is decreasing with increase in temperature.  All three lattice parameters become equal after tetragonal to cubic phase transition ($T_C$ ~763 K) for $x = 0$, and 0.06 composition. However, for $x = 0.09$, $T_C$ increased to ~773 K, in agreement with $c/a$ ratio. Variation of unit cell volume as a function of temperature for $x = 0$, 0.06 and 0.09 is shown in Fig. 2d. The volume was found to decrease with increase in temperature for all measured three samples upto Curie point ($T_C$).

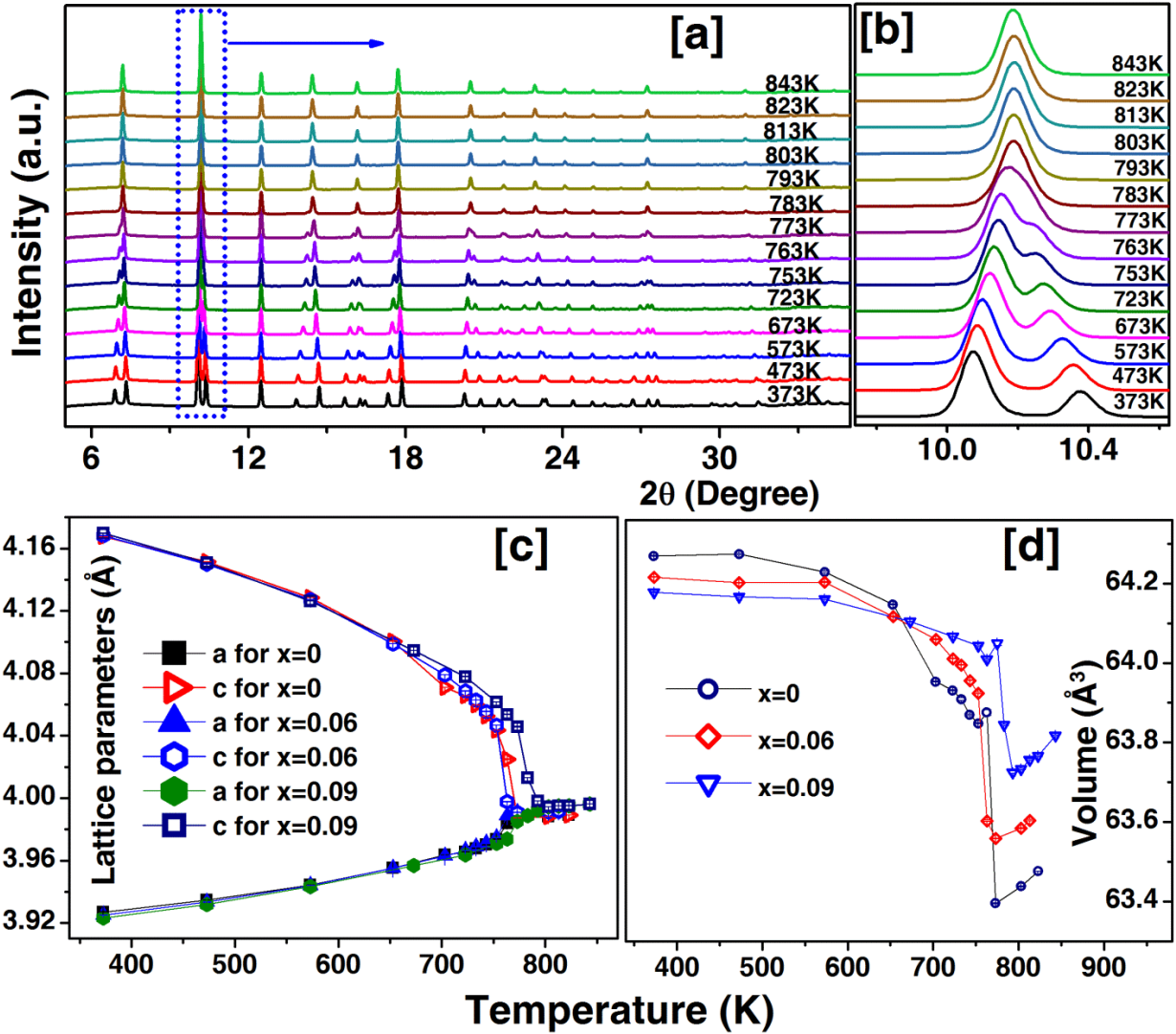

***Figure 2.*** *Temperature-dependent SRPXRD data of $(Pb_{1-x}Bi_x)(Ti_{1-x}Mn_x)O_3$ for $x = 0.09$ composition, where (a) SRPXRD spectra at different temperatures, (b) Highlighted spectra of most intense x-ray peak, (c) Calculated lattice parameters as a function of temperatures for $x = 0, 0.06, 0.09$ compositions, and (d) Calculated unit cell volume as a function of temperatures for $x = 0, 0.06, 0.09$ compositions.*

Below $T_C$, thermal expansion coefficient, $\alpha(T)$, was negative, i.e., unit cell volume contracted with temperature [Fig. 2d] for all three samples. The decrease in unit cell volume increased as T $\rightarrow T_C$. After phase transition temperature, a positive thermal expansion was observed. Hence, structural transition from Ferro to Para can be related to the nature of thermal expansivity of these materials. Thus, tetragonal distortion of ferroelectric phase must be related to negative thermal expansion coefficient. All samples are tetragonal at room temperature, except for $x$ = 0.50. Hence, it can also be expected that for higher Bi/Mn modified PbTiO$_3$ systems with $0.09 \leq x < 0.37$, will have negative thermal expansion behavior similar to $x$ = 0, 0.06 and 0.09 until phase transition temperature. The negative thermal expansion coefficient was calculated from the observed reduction in the volume data as a function of temperature using equation: $\frac{\Delta V}{V_o} = \alpha_v \, \Delta T$, where $V_0$ is the volume of the unit cell at 373 K, $V$ volume of the unit cell as a function of temperature, $\alpha_v$ thermal expansion co-efficient, $\Delta T$ temperature difference, and $\Delta V$ are change in volume. Thermal expansion coefficient was estimated to be -1.571*10$^{-5}$ /K, -2.44*10$^{-5}$ /K, and -5.025*10$^{-6}$ /K for $x$ = 0, 0.06, and 0.09 samples respectively by fitting the estimated volume data as a function of temperature[21, 22].

Correlation between negative thermal expansion and tetragonality of PbTiO$_3$ related structures can be associated to the nature of TiO$_6$ and PbO$_{12}$ polyhedra. In cubic PbTiO$_3$ (T > T$_C$, i.e. 763 K) the TiO$_6$ and PbO$_{12}$ polyhedra are perfectly symmetric. The symmetry in cubic phase is due to equal bond strengths of Pb1-O2 and Pb2-O2 bonds (atoms nomenclature are shown in Fig. S3). As the sample is cooled down from T$_C$, the symmetry of the polyhedra are destroyed as lattice parameters 'c' become larger than 'a' or 'b'. This happens due to the strong hybridization of Pb6s-O2p electrons. This should have resulted in the contraction of the c-axis if both Pb1 and Pb2 would have bonded equivalently to O2. However, the presence of TiO$_6$ octahedra does not allow such contraction and results in O2 being asymmetrically shared between Pb1 and Pb2. Hence, in the tetragonal phase, bond strength of Pb1-O2 is much stronger than Pb2-O2. On the other hand, this powerful Pb1-O2 bond also creates an asymmetry in the TiO$_6$ octahedral, elongating TiO$_6$ octahedra along the c-axis and moving the O-cage upwards. Ti is also displaced due to changes in the position of O2 and O1. Hence, below T$_C$ both polyhedra are distorted. It is interesting to notice that during cooling below T$_C$, rate of increase of 'c' lattice parameter is much more (2~10 times) than that of 'a or b' parameters. This ratio increases with increasing temperature and is strongest near T$_C$. This may be responsible for the negative thermal expansion of these materials. Above T$_C$, thermal energy supplied to the lattice further enlarges all bonds, thereby increasing lattice parameters. Hence, it is interesting to envisage that the same Pb2-O2 bond which was decreasing below T$_C$, increases beyond T$_C$. BiMnO$_3$ modified PbTiO$_3$ are tetragonal up to $x$ = 0.37 [Fig.1a]. Although high-temperature structural studies could not be investigated for all samples, it may be expected that similar behavior may exist for all sample with $x < 0.37$[21, 23-25].

Raman spectroscopy is an excellent tool to assess lattice vibrations, structural distortion and lattice strains associated with the substitution in PBTM samples. Twelve optic modes ($3T_{1u} + T_{2u}$) are observed in the cubic phase, where triply degenerate $T_{1u}$ modes are infrared active and $T_{2u}$ are silent modes which are neither Raman nor infrared active [26]. Tetragonal phase exhibits fifteen Raman active modes. The irreducible representation can be written as $\Gamma_{Raman} = 3A_1(TO) + 3A_1(LO) + B_1 + 4E(TO) + 4E(LO)$.

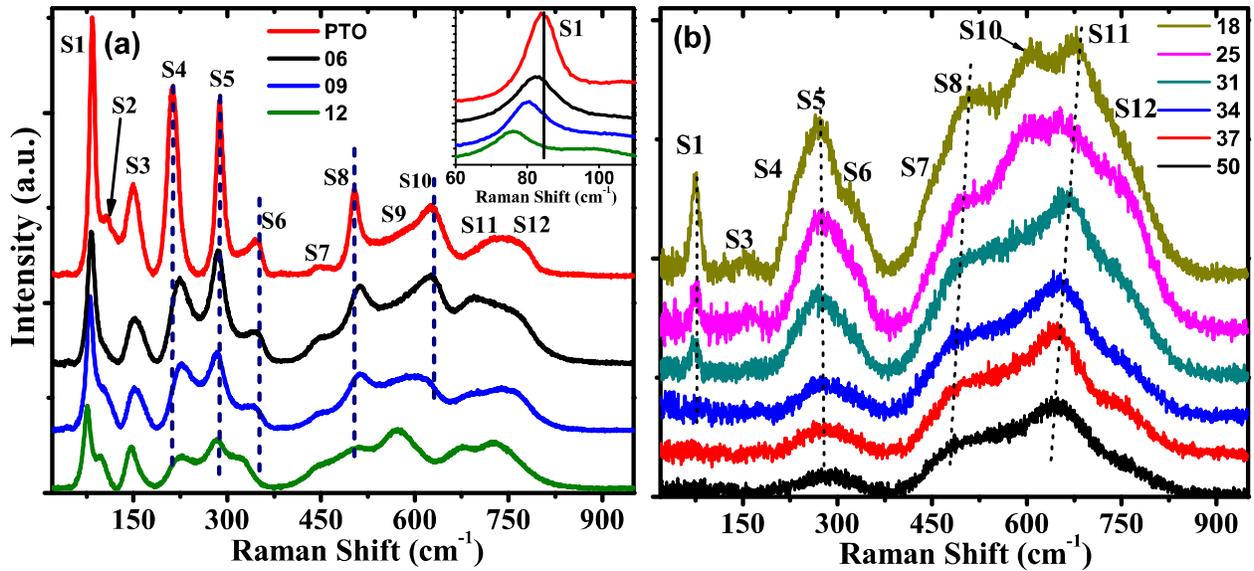

***Figure 3**(a and b): Un-polarised room temperature Raman spectra of (Pb$_{1-x}$Bi$_x$)(Ti$_{1-x}$Mn$_x$)O$_3$ (x = 0, 0.06, 0.09, 0.12, 0.18, 0.25, 0.31, 0.34, 0.37 and 0.50) system. Inset of (a) show the substitution evolution of soft mode S1.*

It is evident from the recorded spectra that there are total 12 distinct modes of vibration for pure PbTiO$_3$ system in the spectral range of 20-950 cm$^{-1}$ (S1 to S12). The observed feature of Raman spectra confirms the tetragonal structure of PbTiO$_3$ under ambient condition [27]. With increasing substitution, a tetragonal to cubic-like phase transformation is witnessed. These modes remain consistent in all compositions with $x \leq 0.12$ (Fig. 3a) and exhibit some changes in mode frequency. However, for $0.18 < x < 0.50$, few modes get merged and disappear from the spectra (Fig. 3b). All the observed phonon modes in Raman spectra for (Pb$_{1-x}$Bi$_x$)(Ti$_{1-x}$Mn$_x$)O$_3$ are tabulated in Table-1.

**Table 1:** The observed phonon modes and their assignment for *(Pb₁₋ₓBiₓ)(Ti₁₋ₓMnₓ)O₃* (*x* = 0, 0.06, 0.09, 0.12, 0.18, 0.25, 0.31, 0.34, 0.37 and 0.50) system. The mode frequencies are in cm⁻¹.

| Phonon Modes | PTBM (*x* = 0) | PTBM (*x*=0.06) | PTBM (*x*=0.09) | PTBM (*x*=0.12) | PTBM (*x*=0.18) | PTBM (*x*=0.25) | PTBM (*x*=0.31) | PTBM (*x*=0.34) | PTBM (*x*=0.37) | PTBM (*x*=0.50) | Mode Assignment |
|---|---|---|---|---|---|---|---|---|---|---|---|
| S1 | 84.2 | 82.5 | 80.5 | 76.4 | 75.4 | 75.8 | 73.7 | - | - | - | E (1TO) |
| S2 | 110.9 | 107.7 | 104.4 | 98.1 | - | - | - | - | - | - | E (1LO) |
| S3 | 149.2 | 151.5 | 152.2 | 149.9 | 158.6 | 163.3 | - | - | - | - | $A_1$ (1TO) |
| S4 | 213 | 224.2 | 226.1 | 226.8 | 233.7 | 237.1 | - | - | - | - | E (2TO) |
| S5 | 287.9 | 284.9 | 283.3 | 282.9 | 274.3 | 274 | 271.4 | 272.7 | 271.9 | 273.6 | $B_1$+E |
| S6 | 346.3 | 348.8 | 345.5 | 322.4 | 323.5 | 320.1 | 315.9 | 313.4 | 311 | 309.2 | $A_1$ (2TO) |
| S7 | 446.2 | 446.1 | 446.1 | 442.5 | - | - | - | - | - | - | E (2LO)+ A1 (2LO) |
| S8 | 503.7 | 509.5 | 508.9 | 506.4 | 507.3 | 496.3 | 492.4 | 489.2 | 487.3 | 482.6 | E (3TO) |
| S9 | 587.6 | 584.4 | 581.4 | - | - | - | - | - | - | - | |
| S10 | 626.8 | 623.3 | 607.1 | 579.9 | 595.7 | 590.8 | 569.9 | 551.2 | 559.2 | 545.9 | $A_1$ (3TO) |
| S11 | 724.7 | 699.7 | 691.9 | 676.4 | 680.9 | 672.3 | 665.1 | 651.9 | 647.7 | 648.4 | E (3LO) |
| S12 | 763.8 | 757.7 | 745.5 | 734.6 | 752.8 | 7515 | 752.8 | 753.1 | 750.7 | 748.3 | $A_1$ (3LO) |

Phonon modes below 150 cm⁻¹ i.e., E(1TO) (~84 cm⁻¹) and $A_1$(TO) (~150 cm⁻¹) are considered as soft phonon modes originating from the $T_{1u}$ (TO) mode of the cubic phase mainly arises due to the vibration of Pb-ion with respect to $TiO_6$ octahedra [28, 29]. A weak E(1LO) i.e., S2 mode vanishes after *x* = 0.12. For *x* > 0.34 all soft modes vanish. All three soft modes are a signature of A-site ordering in the $ABO_3$ structure [30]. Gradual loss of these modes hints at loss of order at the A-site in substituted samples.

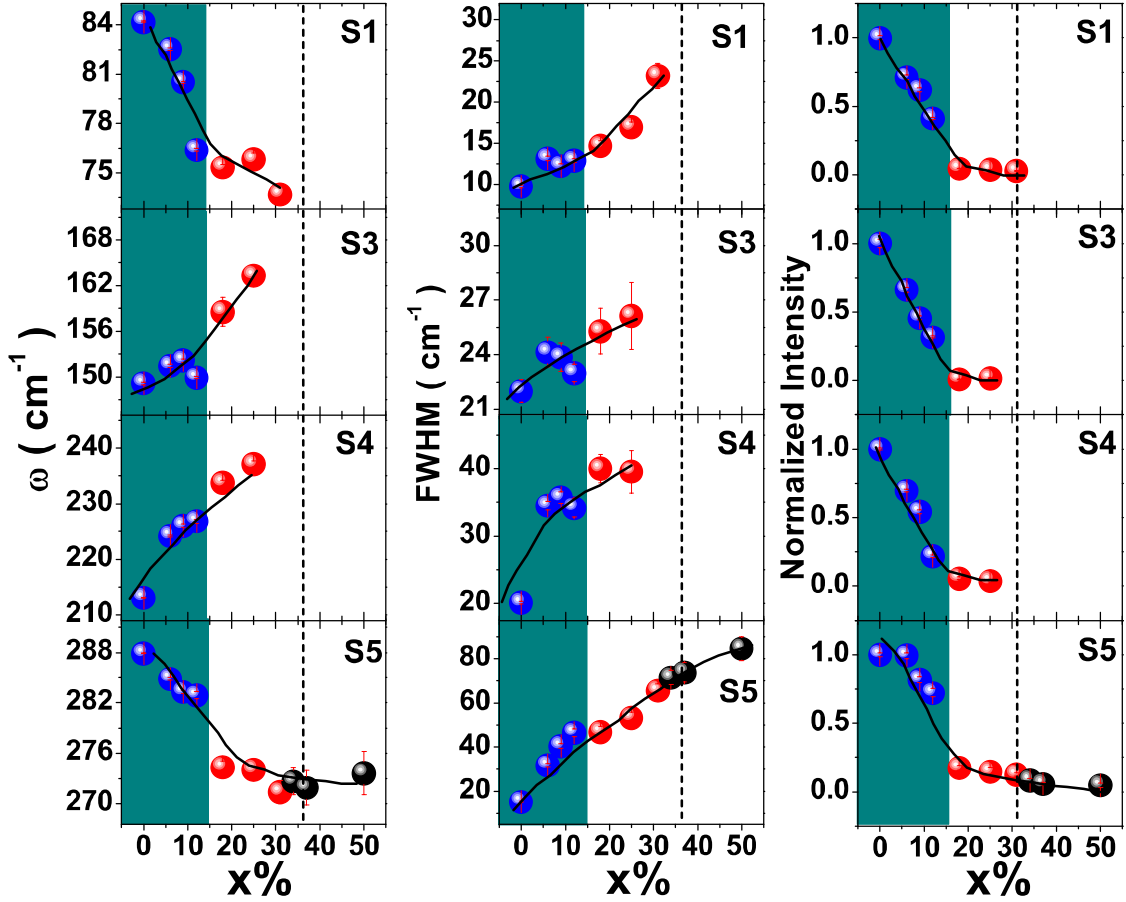

***Figure 4:*** *Substitution dependence of mode frequency, FWHM and normalized intensity for modes S1, S3, S4 and S5 of $(Pb_{1-x}Bi_x)(Ti_{1-x}Mn_x)O_3$ (x = 0, 0.06, 0.09, 0.12, 0.18, 0.25, 0.31, 0.34, 0.37 and 0.50). Shaded region show that maximum changes in the phonon self-energy parameters and integrated intensity are observed till x = 0.12 of substitution. Dotted vertical line shows that dominating phase beyond ~ 37 % of substitution is cubic as all the parameters are nearly invariant of substitution.*

Fig.4 shows variation of mode frequency, full-width at half maximum (FWHM)/damping coefficient and intensity of phonon modes (S1, S3, S4, and S5). Following observations follows: (i) S1: frequency sharply decreases (~9.3 %) for $x \leq 0.12$, marginally decreases (~2.1 %) upto $x = 0.31$; thereafter disappears for $x > 0.31$ (ii) mode S3 and S4 shows large increase in mode frequency ~8 % and 14 %, respectively, with substitution upto $x = 0.25$ and on further increase in $x$ concentration mode S3 disappear and mode S4 merge with mode S5; (iii) frequency of mode S5 shows continues decrease upto $x = 0.12$, and on further increase in substitution, it shows mode softening by ~3.04 % at $x = 0.18$ and thereafter it remains almost constant; (iv) damping coefficient of modes (S1, S3, S4, and S5) shows large increase (as large as 400 %) with increase in concentration clearly suggesting that disorders are increasing linearly with the substitution. As the damping coefficient is inversely proportional to the life time of a phonon mode, a large increase in damping coefficient with increase in substitution suggests that phonon life time decreasing sharply, indirectly reflecting the presence of disorder in the system (v) the intensity of modes (S1, S3, S4, and S5) shows sharp drop on the increase in substitution up to $x = 0.12$ and remain almost constant on further increase in $x$ concentration.

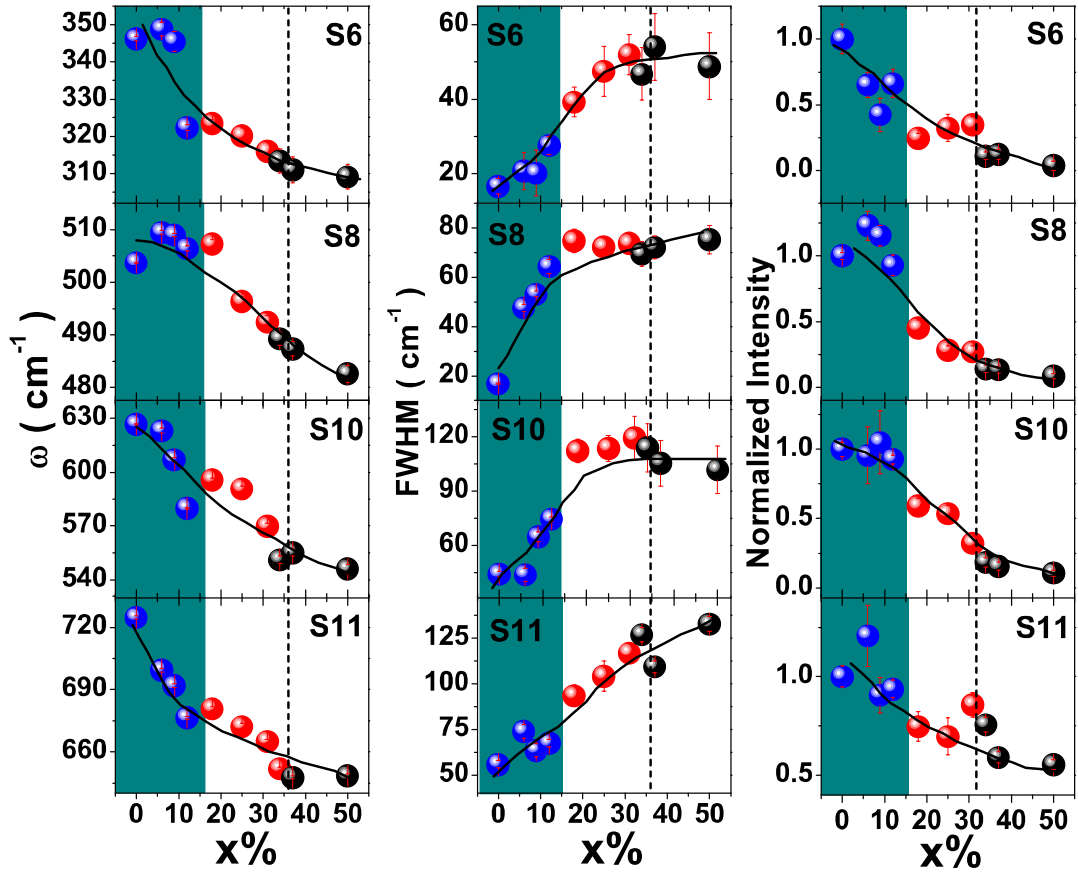

***Figure 5:*** *Substitution dependence of mode frequency, FWHM and normalized intensity for modes S6, S8, S10 and S11 of $(Pb_{1-x}Bi_x)(Ti_{1-x}Mn_x)O_3$ ($x = 0$, 0.06, 0.09, 0.12, 0.18, 0.25, 0.31, 0.34, 0.37 and 0.50). Shaded region show that maximum changes in the phonon self-energy parameters and integrated intensity are observed till $x = 0.12$ of substitution. Dotted vertical line shows that dominating phase beyond ~ 37 % of substitution is cubic as all the parameters are nearly invariant of substitution.*

Based on studies on S6, S8, S10, and S11 (Fig. 5), following observation can be made: (i) frequency of mode S6 remain nearly constant upto $x = 0.09$ and shows mode softening (~6.7 %) at $x = 0.12$, and decreases on further increase in $x$ concentration upto $x = 0.31$, and after that it remains almost constant; (ii) frequency of mode S8 shows slight increase upto $x = 0.12$ and decreases on further increase in $x$ concentration; (iii) modes S10 and S11 shows a large decrease in mode frequency upto $x = 0.34$ and thereafter remain nearly constant with increase in substitution; (iv) damping coefficient of mode S6 and S11 increase upto $x = 0.34$ and remain constant after that, on the other hand damping coefficient of mode S8 and S10 shows increase upto $x = 0.12$ and thereafter it remains almost constant. The intensity of mode S6, S8, S10 and S11 shows a large decrease (~80 %) till $x = 0.31$ and after that remains constant.

It is evident from the above observation that there is a systematic structural distortion in $(Pb_{1-x}Bi_x)(Ti_{1-x}Mn_x)O_3$ with an increase in substitution. The crystal radius of $Pb^{2+}$(XII) (~1.63 Å) is smaller than $Bi^{3+}$(XII) (~1.59 Å). On the other hand, the crystal radius of $Mn^{3+}$(VI) (~0.72 Å for low spin and ~0.785 Å for high spin) is either smaller or larger compared to $Ti^{4+}$(VI) (~0.745 Å) depending on the bond strength of Mn with O ions. As Bi is already smaller than Pb, it is most likely that Mn-O bonds will be longer. Hence, the chance of smaller crystal radius of Mn is lesser than the larger version, thereby resulting in a high spin state. Presence of all phonon modes associated with tetragonality of $ABO_3$ structure confirms a tetragonal $(Pb_{1-x}Bi_x)(Ti_{1-x}Mn_x)O_3$ phase for $x \leq 0.12$. Further substitution shows renormalization of phonon modes self-energy parameters, i.e. mode frequency and damping coefficient; reflecting indications of clear structural distortion. A morphotropic phase boundary in these samples, with co-existent tetragonal and pseudo-cubic phases, [31] may be responsible for such structural distortions. Soft modes in these systems are extremely sensitive to any perturbation, e.g., pressure (external as well as internal), temperature, etc. Such a perturbation in form of substitution is reflected in PBTM samples. With increasing substitution, broadening, merging and finally, the disappearance of phonon modes hint at a tetragonal-cubic phase transition. The tetragonal phase for lower substitution ($x \leq 0.12$), followed by a gradual appearance of cubic phase is evident from Fig. 4 and 5, ultimately changing to a cubic phase after ~ 34 % substitution.

Microstructure analyses of the PBTM samples were performed with FESEM study. Gold-sputtered pellets were used to see the Micrographs of the PBTM samples. Micrographs were collected in the secondary electron mode. Images of all the PBTM samples are shown in Fig.

[14]

6(a-i). Grains are compactly packed in all the samples. Average grain size, estimated using Image J software, decreases from $14.55 \pm 4.82$ μm for $x = 0.06$ to $11.53 \pm 4.32$ μm for $x = 0.50$ and can be related to reduction in the diffusion coefficient of Bi/Mn substitution elements [1, 32].

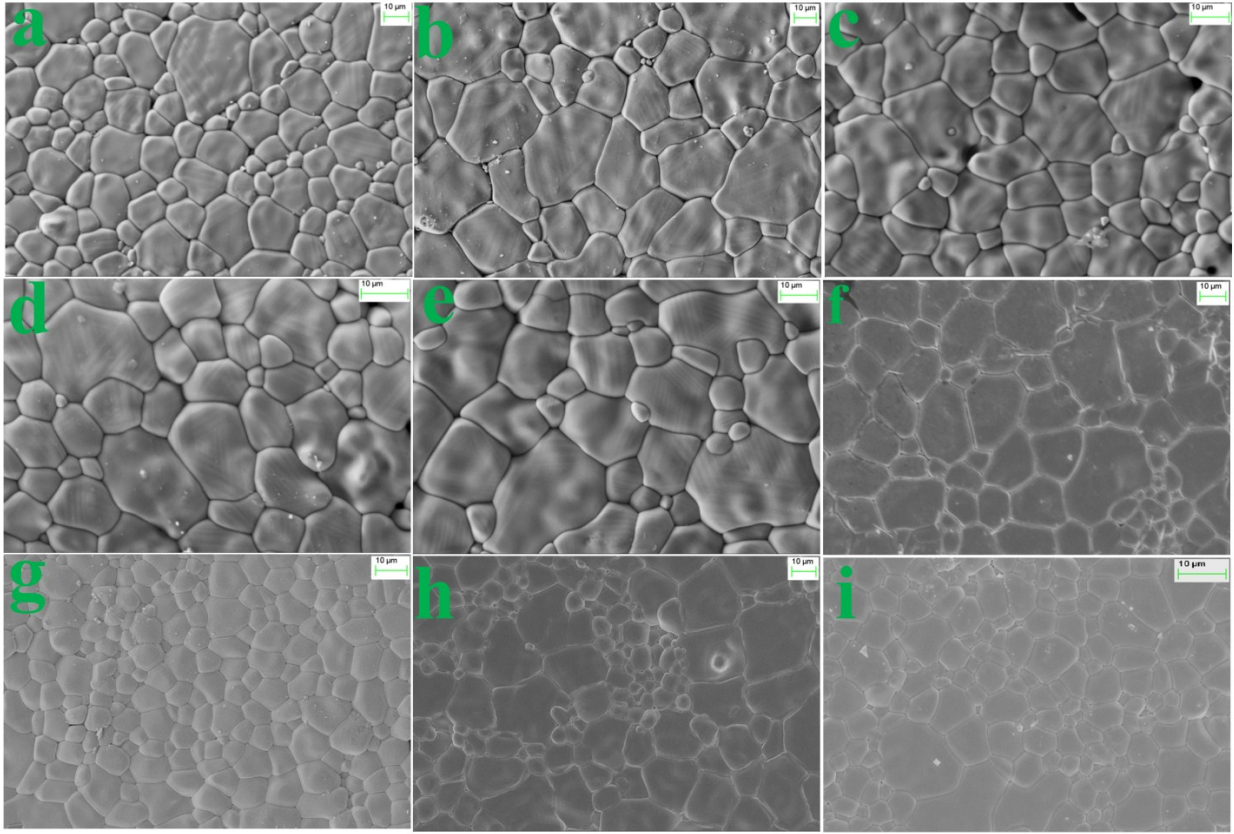

**Figure 6**. *Surface microstructural analyses of $(Pb_{1-x}Bi_x)(Ti_{1-x}Mn_x)O_3$ samples where (a) x = 0.06, (b) x = 0.09, (c) x = 0.12, (d) x = 0.18, (e) x = 0.25, (f) x = 0.31, (g) x = 0.34, (h) x = 0.37, and (i) x = 0.50 compositions.*

To understand the surface morphology of the PBTM samples, The FESEM image and elemental mappings for all the elements are performed. A representative EDS spectrum and elemental mappings along with the secondary electron micrograph for $x = 0.12$ composition are shown in Fig. 7(a-g). Spot EDS was performed on grain and grain boundary regions to confirm chemical composition and homogeneity of the sample. Chemical homogeneity of Pb, Ti, Bi, Mn, and O was also confirmed using area EDS elemental analysis [33, 34]. A combined graph of area spectra [Fig. 7g] provides atomic and the weight percentage of constituent elements [Table in Fig. 7]. The obtained atomic and weight percentage of the composition is well matched with the target sample (Pb$_{0.88}$Bi$_{0.12}$Ti$_{0.88}$Mn$_{0.12}$O$_3$) within the errors related to EDS analysis. Hence, there is no secondary phase perceived in the sample, and all the elements are homogeneously distributed across the sample.

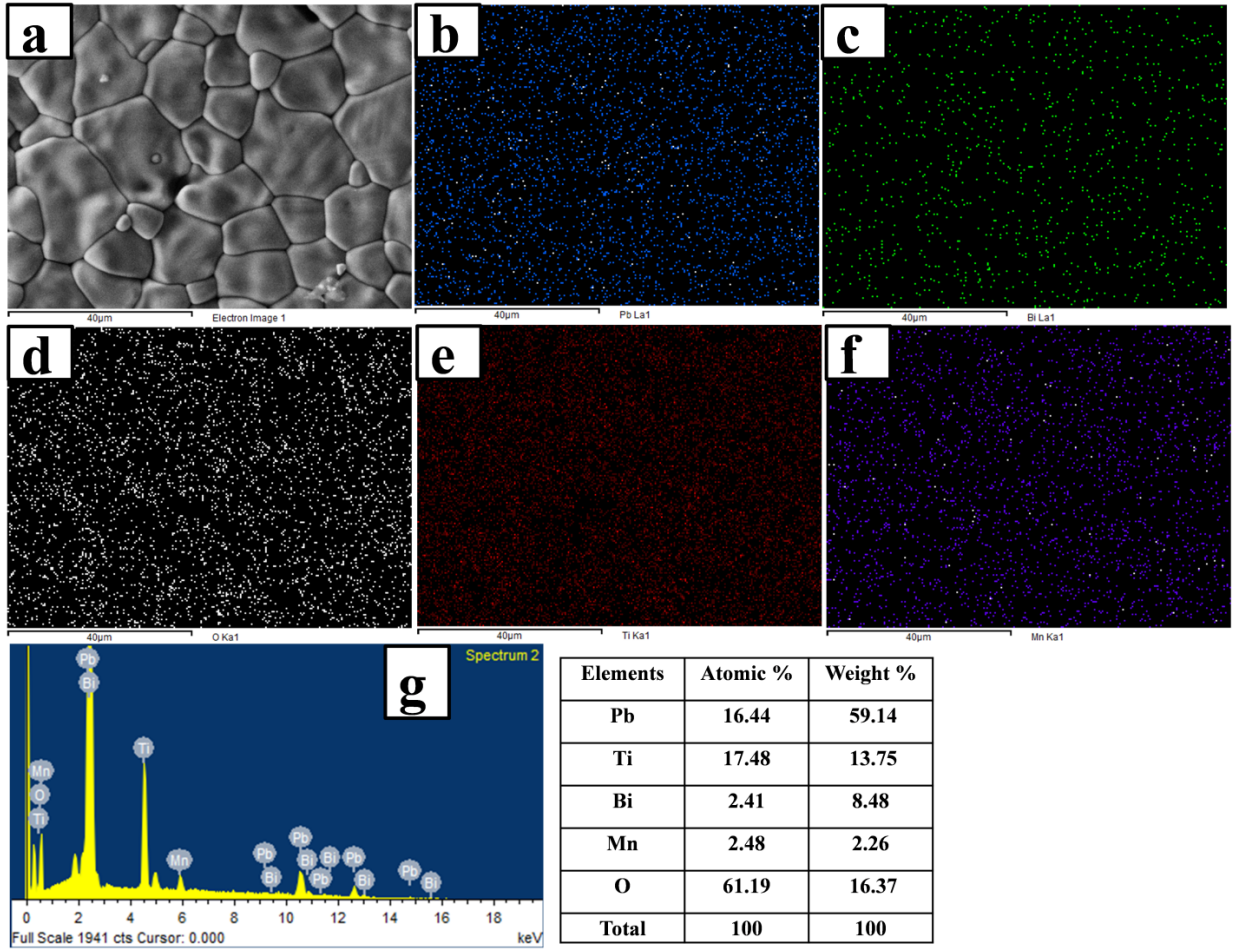

| Elements | Atomic % | Weight % |
|----------|----------|----------|
| Pb | 16.44 | 59.14 |
| Ti | 17.48 | 13.75 |
| Bi | 2.41 | 8.48 |
| Mn | 2.48 | 2.26 |
| O | 61.19 | 16.37 |
| Total | 100 | 100 |

**Figure 7.** *FESEM-EDS elemental mapping of (Pb$_{1-x}$Bi$_x$)(Ti$_{1-x}$Mn$_x$)O$_3$ for x = 0.125 composition where (a) secondary electron image and corresponding elemental mapping of the element (b) Pb, (c) Bi, (d) O, (e) Ti, (f) Mn elements, (g) Area EDS spectrum and right table for the atomic and weight percentage of various elements.*

To explore the physical properties of the PBTM samples, temperature-dependent dielectric constant and loss (tanδ) measurements were carried out at various frequencies. Dielectric constant and tanδ versus temperature for 0.06 ≤ x ≤ 0.25 compositions are shown in the Fig. 8(a-e). For $0.06 \leq x \leq 0.18$, a single dielectric catastrophic anomaly is observed at a certain temperature (Curie point, $T_C$) corresponding to a structural phase transition from polar tetragonal to the centrosymmetric cubic phase of a perovskite structure[35, 36]. $PbTiO_3$ has a sharp transition at $T_C$ ~763 K for ferroelectric to paraelectric phase [37]. For $x = 0.06$ composition, $T_C$ is invariant ~763 K. However, $T_C$ increased to ~773 K for $x = 0.09$ composition which is 10 K temperature more than the pure $PbTiO_3$ system. $T_C$ for $x = 0$, 0.06, and 0.09 was also confirmed from temperature-dependent SRPXRD (Fig. S4 (a-c) and Fig. 2a). $T_C$ reduces to lower temperatures for $x \geq 0.12$ [Fig. 8(c-e)]. For $x = 0.25$ composition, the negative dielectric constant was observed after ~600 K, a phenomenon not possible in capacitive materials. Negative $\varepsilon_r$ may be due to the conductive nature of the material. However, the dielectric anomaly is still evident at $T_C$ ~700 K. In these samples with higher substitution, the dielectric anomaly was not observed.

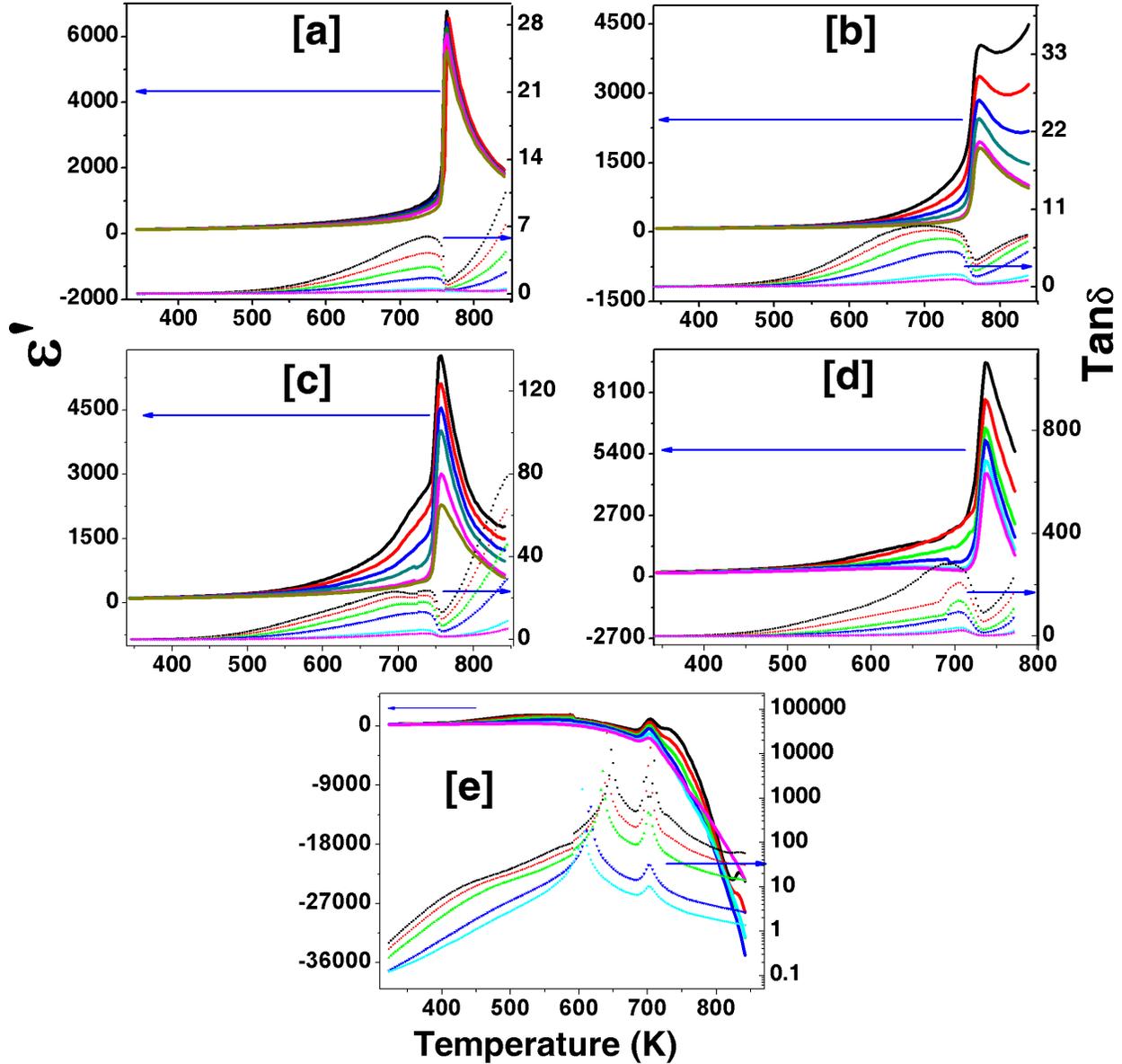

**Figure 8.** *Temperature-dependent dielectric constant and tanδ of the (Pb$_{1-x}$Bi$_x$)(Ti$_{1-x}$Mn$_x$)O$_3$ samples where (a) x = 0.06, (b) x = 0.09, (c) x = 0.12, (d) x = 0.18, and (e) x = 0.25 compositions.*

In the ABO$_3$ structure of PbTiO$_3$, the Pb1-O2 bond strength is a determining factor as has been discussed before. Substitution of Pb by Bi introduces stronger Bi-O2 bonds for the lower amount of substitutions, due to the smaller size of Bi compared to Pb [12]. These stronger Bi-O2 bonds create larger distortion along *c*-axis and are supported by a larger size Mn$^{3+}$. This results in an increase of *c/a* ratio. However, this size effect cannot extend indefinitely. The proximity of

[18]

Mn and O1 ions will soon reach a minimum and after that elongation will be restricted. After that, arrangements of ions happen by enlarging along '*a* or *b*' axes. This restructures electronic clouds around Pb/Bi ions and thereby bonding with O2 is modified reducing *c*-axis. It is observed that *c/a* increases until $x = 0.12$ and after that decreases.

**Conclusions**

Polycrystalline powders of $(Pb_{1-x}Bi_x)(Ti_{1-x}Mn_x)O_3$ ($0 \leq x \leq 0.50$) were successfully prepared via the sol-gel method. Structural properties were investigated with Rietveld refinement of the SRPXRD data and Raman spectroscopy. Tetragonal phase (*P4mm*) was identified for $0 \leq x \leq 0.37$ samples whereas cubic phase (*Pm3m*) for $x = 0.50$ sample. Lattice parameters '*a*' slightly decreases while '*c*' increases upto $x = 0.12$ composition. Hence, tetragonality (*c/a* ratio) nominally increased initially in the range $0 \leq x \leq 0.12$ from 1.064 to ~1.066, thereafter decreasing continuously in the range $0.12 \leq x \leq 0.50$ to 1, for $x = 0.5$ (*c/a* = 1). A Ferro-para phase transition in these materials can be related to tetragonal to cubic phase transition using composition dependent dielectric studies of the materials. Synchrotron-based temperature-dependent SRPXRD was carried out for $x = 0$, 0.06, and 0.09 samples to study the lattice dynamics of the phase transition temperature. Lattice parameters '*c*' decreased while '*a*' increased with increase in temperature. The phase transition temperature, $T_C$, increased from 763 K in $x \leq 0.06$ to 773 K in $x = 0.09$ samples. Thermal expansion coefficient was obtained to be -$1.571*10^{-5}$ /K, $-2.44*10^{-5}$ /K, and $-5.025*10^{-6}$ /K for $x = 0$, 0.06, and 0.09 samples respectively by fitting the estimated volume data as a function of temperature. Dielectric properties also reveal same phase transition temperature. Unit cell volume was found to decrease with increase in temperature until $T_C$, showing negative thermal expansion.

**Acknowledgements**

One of the authors is thankful to University Grants Commission for providing research fellowship (NFO – 2015 – 17 – OBC – UTT - 28455). Principle investigator expresses sincere thanks to Indian Institute of Technology Indore for funding the research. The authors sincerely thank Sophisticated Instrument Centre (IIT Indore) for FESEM studies. Pradeep Kumar and Sunil Kumar thanks DST-India for the award of research grant under INSPIRE Faculty scheme and AMRC, IIT Mandi for the Raman facility. One of the authors (Sajal Biring) acknowledges the financial support from Ministry of Science and Technology, Taiwan (MOST 105 − 2218 − E − 131 - 003 and 106 − 2221 − E − 131 - 027).

**Supplementary Information**

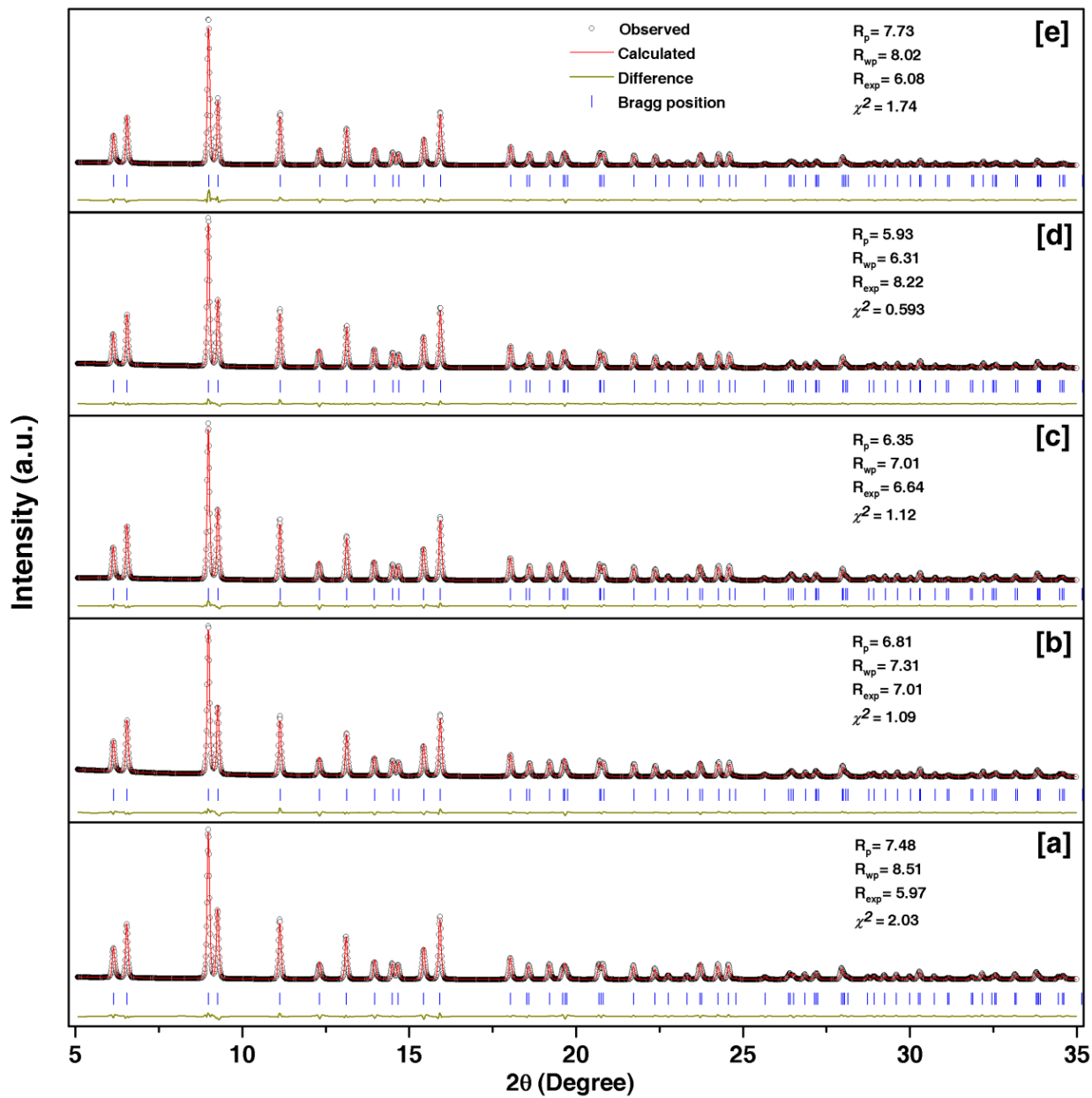

**Figure S1**. Rietveld refinement goodness of fitting for Pb$_{(1-x)}$Bi$_x$Ti$_{(1-x)}$Mn$_x$O$_3$ samples where (a) $x = 0$, (b) $x = 0.06$, (c) $x = 0.09$, (d) $x = 0.12$, and (d) $x = 0.18$ compositions. Fitting parameters

are provided in the corresponding figure. The hollow black symbols are observed data, red solid line calculated data, below dark yellow solid line difference of observed and calculated data, and vertical blue bar lines are Bragg positions with the corresponding *P4mm* space group.

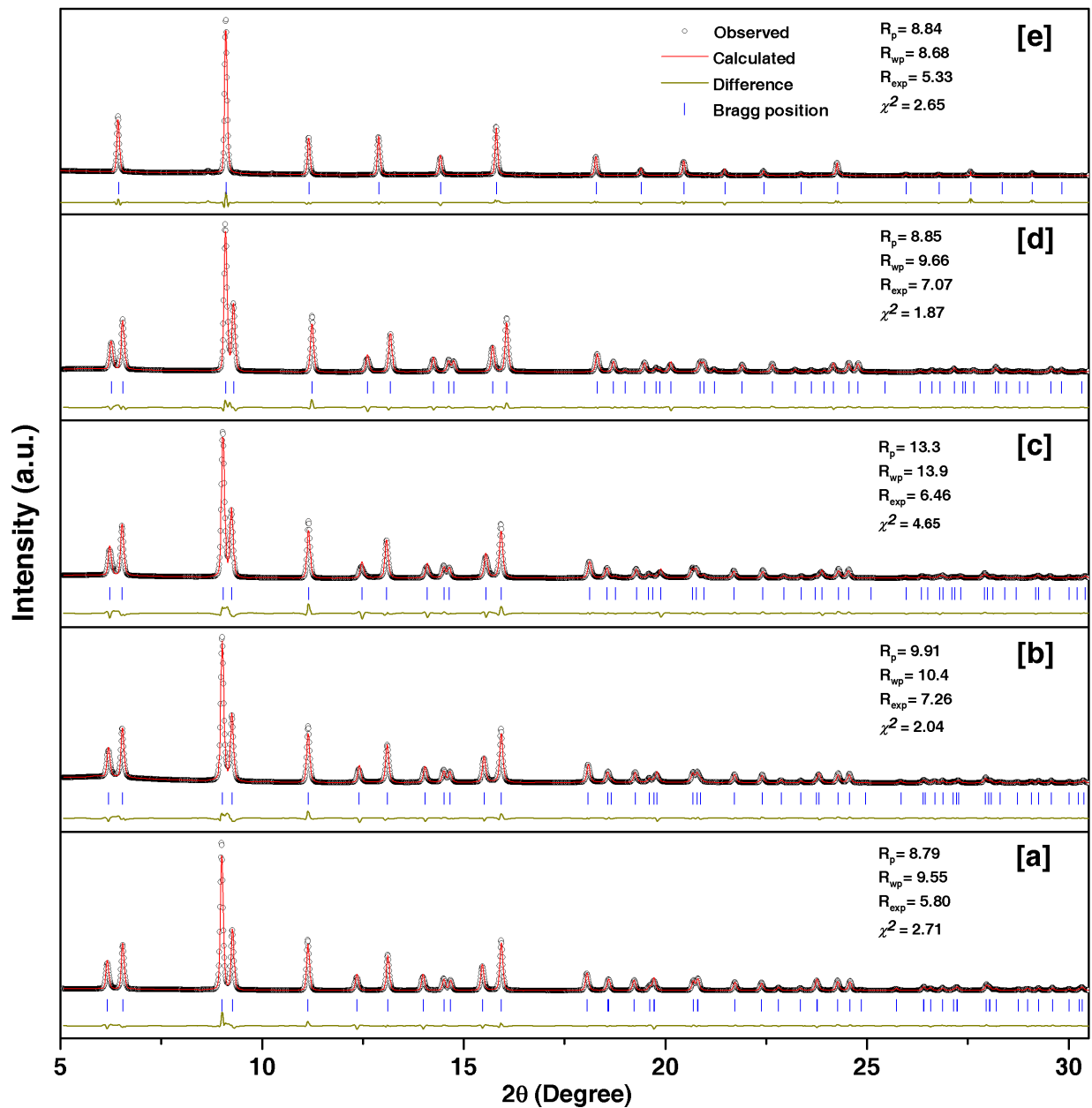

**Figure S2**. Rietveld refinement goodness of fitting for $Pb_{(1-x)}Bi_xTi_{(1-x)}Mn_xO_3$ samples where (a) $x = 0.25$, (b) $x = 0.31$, (c) $x = 0.34$, (d) $x = 0.37$, and (e) $x = 0.50$ compositions. Fitting parameters are provided in the corresponding figure. The hollow black symbols are observed data, red solid

line calculated data, below dark yellow solid line difference of observed and calculated data, and vertical blue bar lines are Bragg positions with the corresponding space group.

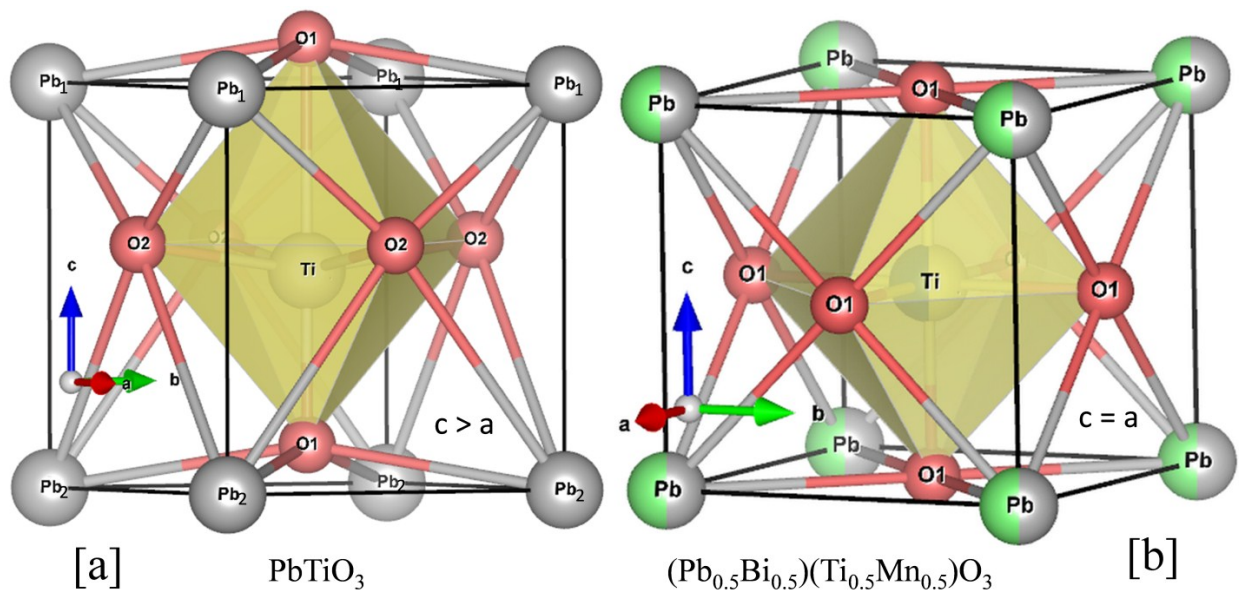

**Figure S3**. Unit cell structure after Rietveld refinement (a) Pure $PbTiO_3$ with tetragonal structure, (b) $Pb_{(1-x)}Bi_xTi_{(1-x)}Mn_xO_3$ for $x = 0.50$ composition with cubic structure. Mark of Pb with 1 and 2 is just for explanation.

**Table S1.** Position parameters of the $Pb_{(1-x)}Bi_xTi_{(1-x)}Mn_xO_3$ $(0 \leq x \leq 0.50)$ range samples after the Rietveld refinement of SRPXRD data.

| Atoms | Coordinates | Composition ($x$) | | | | | | | | | |
|---|---|---|---|---|---|---|---|---|---|---|---|
| | | **0.00** | **0.06** | **0.09** | **0.12** | **0.18** | **0.25** | **0.31** | **0.34** | **0.37** | **0.50** |
| **Pb/ Bi** | $x$ | 0 | 0 | 0 | 0 | 0 | 0 | 0 | 0 | 0 | 0 |
| | $y$ | 0 | 0 | 0 | 0 | 0 | 0 | 0 | 0 | 0 | 0 |
| | $z$ | 0 | 0 | 0 | 0 | 0 | 0 | 0 | 0 | 0 | 0 |
| **Ti/ Mn** | $x$ | 0.5 | 0.5 | 0.5 | 0.5 | 0.5 | 0.5 | 0.5 | 0.5 | 0.5 | 0.5 |
| | $y$ | 0.5 | 0.5 | 0.5 | 0.5 | 0.5 | 0.5 | 0.5 | 0.5 | 0.5 | 0.5 |
| | $z$ | 0.53816 (1) | 0.54018 (1) | 0.53986 (1) | 0.54217 (75) | 0.54204 (1) | 0.54291( 1) | 0.54419 | 0.54363(1) | 0.53907(1) | 0.5 |
| **O₁** | $x$ | 0.5 | 0.5 | 0.5 | 0.5 | 0.5 | 0.5 | 0.5 | 0.5 | 0.5 | 0.5 |
| | $y$ | 0.5 | 0.5 | 0.5 | 0.5 | 0.5 | 0.5 | 0.5 | 0.5 | 0.5 | 0.5 |
| | z | 0.10861 (1) | 0.10844 (1) | 0.10902 (1) | 0.10477 (207) | 0.1180 (1) | 0.11945( 1) | 0.10646 | 0.10311(1) | 0.09403(1) | 0 |
| $O_2$ | x | 0.5 | 0.5 | 0.5 | 0.5 | 0.5 | 0.50 | 0.5 | 0.5 | 0.5 | 0.5 |
| | y | 0 | 0 | 0 | 0 | 0 | 0 | 0 | 0 | 0 | 0 |
| | z | 0.60843 (1) | 0.60615 (0) | 0.60778 (1) | 0.60728 (126) | 0.60445 (1) | 0.60566( 1) | 0.59998(1) | 0.59978(1) | 0.59481(1) | 0.5 |

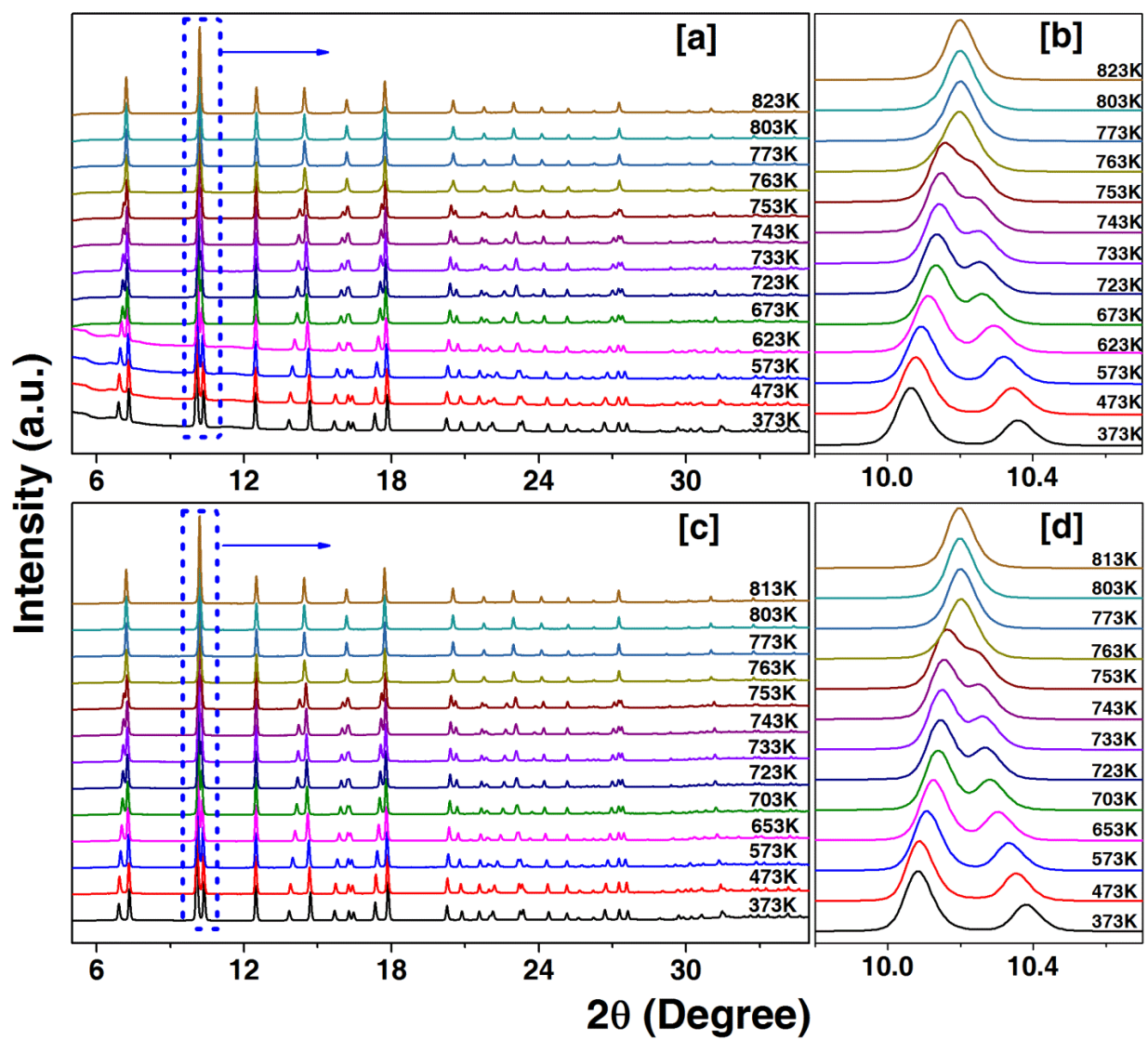

**Figure S4.** Temperature-dependent SRPXRD data of $(Pb_{1-x}Bi_x)(Ti_{1-x}Mn_x)O_3$ for $x = 0$ and $0.06$ composition, where (a) SRPXRD spectra at different temperatures for $x = 0$ composition, (b)

Highlighted spectra of most intense x-ray peak, (c) x-ray spectra at different temperature for $x =$ 0.06 composition, and (d) Highlighted spectra of most intense peak.